# Torsional Weyl-Dirac Electrodynamics

## Mark Israelit[1]

ABSTRACT
-------------------------------------------------------------------------------------------------
*Issuing from a geometry with nonmetricity and torsion we build up a generalized classical electrodynamics. This geometrically founded theory is coordinate covariant, as well gauge covariant in the Weyl sense. Photons having arbitrary mass, intrinsic magnetic currents (magnetic monopoles), and electric currents exist in this framework. The field equations, and the equations of motion of charged (either electrically, or magnetically) particles are derived from an action principle. It is shown that the interaction between magnetic monopoles is transmitted by massive photons. On the other hand the photon is massive only in presence of magnetic currents. We obtained a static spherically symmetric solution, describing either the Reissner-Nordstrom metric of an electric monopole, or the metric and field of a magnetic monopole. The latter must be massive. In absence of torsion and in the Einstein gauge one obtains the Einstein-Maxwell theory.*
-------------------------------------------------------------------------------------------------

## 1. INTRODUCTION

Maxwell's classical electrodynamics is one of the main pillars of theoretical physics. Its perfection and intrinsic completeness have filled us with deep admiration for more than 120 years. There are, however, two crucial electrodynamical phenomena, which are standing outside of this framework**:** magnetic monopoles, and massive photons.

The hypothesis of the magnetic monopole was introduced in 1931 by Paul Dirac [1].

Later he succeeded in setting up a more general theory [2], which led to a dynamical justification of his famous quantization condition. It is worth noting that this condition may be derived semiclassically, or even classically (cf. J.D.Jackson's remarkable textbook [3], as well [4-6]). In 1965 Golhaber [7] derived Dirac's quantization condition from rotational invariance. In his paper was also pointed out the connection between magnetic monopoles and the Aharonov-Bohm effect [8] (cf. also [31], and [9] ).

The idea of massive photons was discussed during the last 50 years by many physicists, beginning with Louis de Broglie [10] in 1934. Bondy and Lyttleton [11] linked massive photons with the cosmological constant $\Lambda$. In a modern interpretation their affirmation might be regarded as a contribution of the photon mass to cosmic dark matter. There exists also a speed-of-light catastrophe in nature that was elegantly displayed by Ardavan [12] some years ago. Massive photons may help to avoid it. Not only in classical theories, but also in quantum field theories a zero-mass photon leads to difficulties and contradictions. Coleman and Weinberg [13] claim that the photon acquires a mass as a result of radioactive corrections. Estimations of the photon mass may be found in a comprehensive review article by Goldhaber and Nieto [14], as well in a recent review [15]. According to the latter paper $m_\gamma \leq 5.34 \times 10^{-60} \text{g} \doteq 3 \times 10^{-27} \text{eV}$.

[1] Dept. of Physics, University of Haifa - Oranim, Tivon-36006, Israel.
E-mail: israelit@macam.ac.il, israelit@physics.technion.ac.il

The purpose of the present work is to consider a geometrically based generalized electrodynamics admitting massive photons and possessing intrinsic electric, as well as magnetic currents (magnetic monopoles). Some basic elements of the formal framework were set up by the present writer in two previous papers [16], [17].

For a moment let us recall the ordinary Maxwell equations:

$$\nabla \cdot \vec{E} = 4\pi \rho, \quad \nabla \times \vec{H} = 4\pi \vec{j} + \frac{\partial \vec{E}}{\partial t}. \tag{1}$$

and

$$\nabla \times \vec{E} = -\frac{\partial \vec{H}}{\partial t}, \quad \nabla \cdot \vec{H} = 0. \tag{2}$$

One sees that magnetic sources are not appearing in this system, so that one has an asymmetry regarding to electric and magnetic currents. This asymmetry becomes manifested better if one rewrites the system (1), (2) in a covariant form:

$$F^{\mu\nu}_{;\nu} = 4\pi J^\mu, \tag{3}$$

and

$$\tilde{F}^{\mu\nu}_{;\nu} = 0. \tag{4}$$

with the dual field tensor given by

$$\tilde{F}^{\mu\nu} = -\frac{\varepsilon^{\mu\nu\alpha\beta}}{2\sqrt{-g}} F_{\alpha\beta}, \tag{5}$$

($\varepsilon^{\mu\nu\alpha\beta}$ stands for the completely antisymmetric Levi-Civita symbol, and $\varepsilon^{0123} = 1$.) It follows from (4) that the field strength tensor $F_{\mu\nu}$ may be expressed in terms of a potential vector $w_\mu$ as

$$F_{\mu\nu} = w_{\mu;\nu} - w_{\nu;\mu}. \tag{6}$$

From (6) one obtains the Maxwell gauge invariance, and turning to the Lorentz gauge one can rewrite (3) in current-free regions as

$$w^\mu_{;\nu;\nu} + w^\nu R^\mu_\nu = 0, \tag{7}$$

with $R^\mu_\nu$ being the Ricci tensor. If the curvature is negligible, one is left with an equation describing a massless particle, having spin 1, the ordinary photon.

Considering equations (3) - (7), one concludes that in a modified electrodynamics including magnetic monopoles the dual field tensor $\tilde\Phi^{\mu\nu}$ must have a non-vanishing divergence, while in a theory admitting massive photons the Maxwell equations must be replaced by Proca-type equations. It is desirable to derive the field equations and the conservation laws from an action principle. Further, as the Proca terms break the Maxwell gauge invariance, one must turn to another (more general) gauge transformation, giving in the limiting case the Maxwell one. Finally the theory must be in accordance with Einstein's general relativity theory, and it has to be based on a suitable geometric framework.

## 2. THE GEOMETRY

An appropriate for our purpose background is Weyl's geometry with torsion. Weyl [18] proposed a generalization of the Riemann geometry soon after Einstein [19] developed the General Relativity Theory, in which gravitation is described in terms of Riemann's geometry. From his geometry Weyl built up a more general theory in which electromagnetism is also described geometrically. However this theory had some unsatisfactory features and did not gain general acceptance. In 1973 Dirac [20] returned to Weyl's theory, but introduced modifications which removed the earlier difficulties. Later Rosen [21] developed the Weyl-Dirac theory and gave a detailed discussion on its physical contents. There were also built up geometrically based theories of gravitation and electromagnetism [22], [23] from the Weyl-Dirac approach.

The original Weyl-Dirac framework is a torsionless geometry with non-metricity. We will generalize this geometry by incorporating torsion. Let us assume that in each point of the 4-dimensional space-time are given a symmetric metric tensor $g_{\mu\nu} = g_{\nu\mu}$, a Weyl connection vector $w_\mu$, the Dirac gauge function $\beta$, as in the original Weyl-Dirac theory [18], [20], [21], and also a torsion tensor $\Gamma^\lambda_{[\mu\nu]}$. In this case the (asymmetric) connection may be written (cf. [24]) as

$$\Gamma^\lambda_{\mu\nu} = \left\{{}^\lambda_{\mu\nu}\right\} + g_{\mu\nu} w^\lambda - \delta^\lambda_\mu w_\nu - \delta^\lambda_\nu w_\mu + C^\lambda_{\mu\nu}, \qquad (8)$$

where $\left\{{}^\lambda_{\mu\nu}\right\}$ is the usual Christoffel symbol with respect to $g_{\mu\nu}$, and the contorsion tensor $C^\lambda_{\mu\nu}$ is given as

$$C^\lambda_{\mu\nu} = \Gamma^\lambda_{[\mu\nu]} - g^{\lambda\beta} g_{\sigma\mu} \Gamma^\sigma_{[\beta\nu]} - g^{\lambda\beta} g_{\sigma\nu} \Gamma^\sigma_{[\beta\mu]}. \qquad (9)$$

We have now a Weyl geometry with torsion, in which both, the direction, and the length of a vector will change in the process of parallel displacement. For a vector $B^\mu$, having the length $B$, defined by $B^2 = g_{\mu\nu} B^\mu B^\nu$, these changes are given by

$$dB^\mu = -B^\sigma \Gamma^\mu_{\sigma\nu} dx^\nu, \qquad dB = B w_\nu dx^\nu. \qquad (10)$$

According to (10) after a parallel displacement around an infinitesimal parallelogram the length of a vector changes by

$$\Delta B = -B W_{\mu\nu} dx^\mu \delta x^\nu, \qquad (11)$$

with the Weylian length curvature tensor

$$W_{\mu\nu} = w_{\mu;\nu} - w_{\nu;\mu}, \qquad (12)$$

and where a semicolon (;) stands for the covariant derivative formed with $\left\{{\lambda \atop \mu\nu}\right\}$. According to (11) the length is not integrable, so that one has an arbitrary standard of length (or gauge) at each point, and local gauge transformations exist

$$B \to \overline{B} = e^\lambda B, \qquad (13)$$

with $\lambda(x^\mu)$ being an arbitrary function of the coordinates. Under transformations (13) the metric tensor, the Weyl connection vector and the Dirac gauge function change as follows (cf. [18], [20], [21]):

$$g_{\mu\nu} \to \overline{g}_{\mu\nu} = e^{2\lambda} g_{\mu\nu}, \quad w_\mu \to \overline{w}_\mu = w_\mu + \lambda_{,\mu}, \quad \beta \to \overline{\beta} = e^{-\lambda}\beta. \qquad (14)$$

It will be assumed that the torsion tensor is gauge invariant, so that in addition to (14)

$$\Gamma^\lambda_{[\mu\nu]} \to \overline{\Gamma}^\lambda_{[\mu\nu]} = \Gamma^\lambda_{[\mu\nu]}. \qquad (15)$$

Relation (12) and the second of transformation (14) led Weyl to interpret $w_\mu$ as the potential vector of the electromagnetic field. In the geometry discussed here one can make use of several curvature tensors. One has the Weylian length tensor (12); further considering a parallel displacement of $B_\mu$ around an infinitesimal parallelogram one obtains the curvature tensor

$$K^\lambda_{\mu\nu\sigma} = -\Gamma^\lambda_{\mu\nu,\sigma} + \Gamma^\lambda_{\mu\sigma,\nu} - \Gamma^\alpha_{\mu\nu}\Gamma^\lambda_{\alpha\sigma} + \Gamma^\alpha_{\mu\sigma}\Gamma^\lambda_{\alpha\nu}. \qquad (16)$$

where a comma denotes partial differentiation. In addition one can consider the second derivative of a vector $B_\mu$ with respect to the connection $\Gamma^\lambda_{\mu\nu}$ defined by (8):

$$B_{\mu:\nu:\sigma} - B_{\mu:\sigma:\nu} = B_\lambda K^\lambda_{\mu\nu\sigma} - 2 B_{\mu:\alpha} \Gamma^\alpha_{[\nu\sigma]}. \qquad (17)$$

The second term on the right side of (17) describes geometric properties invoked by torsion. In order to include these properties in the Lagrangian, one has to replace the arbitrary vector $B_\mu$ by a fundamental geometric quantity that is already present in the framework. For this purpose let us take the tensor $W_{\mu\nu}$ given by (12), so that the additional curvature term will be written as

$$aW_{\mu\lambda:\alpha}\Gamma^{\alpha}{}_{[\nu\sigma]} \quad , \tag{18}$$

with $a$ being an arbitrary constant. In the next section we will make use of (12), (16), and (18) for building up the Lagrangian density.

## 3. THE ACTION PRINCIPLE

Following Dirac we start from an action principle

$$\delta I = \delta \int L \left(-g\right)^{1/2} d^4 x \tag{19}$$

We want to have a theory, in which the equations are both, coordinate covariant, and gauge invariant. Thus the action $I$ must be an invariant with respect to both kinds of transformations (it must be an in-invariant). For a moment let us go back to the original Weyl geometry [18] without torsion, and let us recall Dirac's procedure [20]. Dirac had the following dynamical variables: $g_{\mu\nu}$, $w_\mu$, and $\beta$. From the length curvature tensor (12) he built the scalar $W_{\mu\nu}W^{\mu\nu}$, and from (16), but with $C^\lambda_{\mu\nu} = 0$ (cf. (8)), he formed the scalar $K = g^{\mu\nu}K^\lambda_{\mu\nu\lambda}$. With these two scalars the in-invariant action was given by [20]

$$I_{Dirac} = \int \left[W_{\mu\nu}W^{\mu\nu} - \beta^2 K + k\left(\beta_{,\underline{\mu}} + \beta w^\mu\right)\left(\beta_{,\mu} + \beta w_\mu\right) + 2\Lambda\beta^4 + L_m\right]\left(-g\right)^{1/2} d^4 x \quad . \tag{20}$$

In the action (20) $k$ is an arbitrary parameter, $\Lambda$ is the cosmological constant, $L_m$ stands for the Lagrangian density of matter, and an underlined index is to be raised with $g^{\mu\nu}$. In order to avoid Proca-type terms in the field equations, Dirac took $k = 6$. As a result he obtained a geometrically based generalized representation of gravitation and electromagnetism. There are two gauge freedoms in Dirac's version: that of Maxwell, and the other, given by Weyl's gauge transformations (13), (14). Setting $\beta = 1$ yields the Einstein-Maxwell theory. For this reason the gauge $\beta = 1$ was named [21] the Einstein gauge.

In our modification of the Weyl-Dirac theory the connection (8) is asymmetric, so that the curvature (16), and hence the scalar $K$, will include torsion terms. Further we add to Dirac's Lagrangian density the term

$$aW_{\mu\nu;\alpha}\Gamma^{\alpha}{}_{[\lambda\sigma]} g^{\mu\lambda} g^{\nu\sigma} \quad , \tag{21}$$

stemming from (18) but with $\Gamma$-differentiation (:) replaced by the Riemannian one (;). Finally, the parameter $k$ is not fixed by Dirac's choice; now it may be chosen arbitrarily, and for $k \neq 0$ we obtain a Proca equation [25], instead of Maxwell's one. Moreover, as pointed out by Rosen [21], for values $k < 6$ the Proca field, from the standpoint of quantum mechanics, may be interpreted as an ensemble of bosons, particles of spin 1, and finite mass. Below these particles will be identified with massive photons.

Expressing the curvature scalar K explicitly in terms of $\{^{\lambda}_{\mu\nu}\}$, $\Gamma^{\lambda}_{[\mu\nu]}$, $w_{\mu}$, (cf. (8), (9)) we get the following action integral:

$$I = \int [W^{\lambda\sigma}W_{\lambda\sigma} - \beta^2 R + \beta^2(k-6)w^{\sigma}w_{\sigma} + 2(k-6)\beta w^{\sigma}\beta_{,\sigma} + k\beta_{,\sigma}\beta_{,\underline{\sigma}}$$
$$+ 8\beta\beta_{,\underline{\lambda}}\Gamma^{\sigma}_{[\lambda\sigma]} + \beta^2 \left(2\Gamma^{\alpha}_{[\mu\lambda]}\Gamma^{\lambda}_{[\mu\alpha]} - 4\Gamma^{\alpha}_{[\sigma\alpha]}\Gamma^{\omega}_{[\sigma\omega]} + \Gamma^{\alpha}_{[\mu\lambda]}\Gamma^{\omega}_{[\underline{\mu}\underline{\lambda}]}g_{\alpha\omega} + 8\Gamma^{\alpha}_{[\sigma\alpha]}w^{\sigma}\right) \quad (22)$$
$$+ aW_{\underline{\mu\nu};\alpha}\Gamma^{\alpha}_{[\mu\nu]} + 2\Lambda\beta^4 + L_m \, ](-g)^{\frac{1}{2}} dx^4,$$

with $R$ being the Riemann curvature scalar. In (22) the dynamical variables are: $\Gamma^{\lambda}_{[\mu\nu]}$, $w_{\mu}$, $g_{\mu\nu}$, and $\beta$.

Varying in (22) $w_{\mu}$, and $\Gamma^{\lambda}_{[\mu\nu]}$, making use of symmetries, and differentiating, one can readily prove that in order to get non-contradictory results, one must take for the constant appearing in (22) $a = 4$,. Below this value will be adopted. In the next section the field equations will be derived from the modified in-invariant action integral (22).

## 4. FIELD EQUATIONS AND CONSERVATION LAWS

Varying in (22) the Weyl connection vector $w_{\mu}$, one obtains from (19) the following field equation

$$\left(W^{\mu\nu} - 2\Gamma^{\alpha}_{[\underline{\mu\nu}];\alpha}\right)_{;\nu} = (\tfrac{1}{2})\beta^2(k-6)W^{\mu} + 2\beta^2\Gamma^{\alpha}_{[\underline{\mu}\alpha]} + 4\pi J^{\mu}, \quad (23)$$

where $W_{\mu}$ stands for the gauge-invariant (cf. (14)) Weyl connection vector

$$W_{\mu} = w_{\mu} + (\ln \beta)_{,\mu}, \quad (24)$$

and

$$16\pi J^{\mu} = \frac{\delta L_m}{\delta w_{\mu}}. \quad (25)$$

Considering in the action the variation with respect to $\Gamma^{\lambda}_{[\mu\nu]}$ one has the equation

$$W^{\mu\nu}_{;\lambda} = \beta^2\left(\delta^{\mu}_{\lambda}W^{\nu} - \delta^{\nu}_{\lambda}W^{\mu}\right) + \beta^2\left(\delta^{\nu}_{\lambda}g^{\mu\sigma} - \delta^{\mu}_{\lambda}g^{\nu\sigma}\right)\Gamma^{\alpha}_{[\sigma\alpha]}$$
$$+ (\tfrac{1}{2})\beta^2\left(g^{\sigma\nu}\delta^{\mu}_{\omega}\delta^{\rho}_{\lambda} - g^{\sigma\mu}\delta^{\nu}_{\omega}\delta^{\rho}_{\lambda} - g^{\rho\nu}g^{\sigma\mu}g_{\lambda\omega}\right)\Gamma^{\omega}_{[\sigma\rho]} - 4\pi\Omega^{[\mu\nu]}_{\lambda}, \quad (26)$$

where

$$16\pi\Omega^{[\mu\nu]}_{\lambda} = \frac{\delta L_m}{\delta\Gamma^{\lambda}_{[\mu\nu]}}. \quad (27)$$

Contracting (26) gives:

$$W^{\mu\nu}_{;\nu} = -3\beta^2 W^{\mu} + 2\beta^2\Gamma^{\nu}_{[\underline{\mu\nu}]} - 4\pi\Omega^{[\mu\nu]}_{\nu}. \quad (28)$$

Below we shall regard $J^\mu$ as the electric current density vector, while the magnetic current density vector will be expressed in terms of $\Omega_\lambda^{[\mu\nu]}$. Therefore it is interesting to write down two conservation laws, which follows from (23), and (28)

$$(k-6)\left(\beta^2 W^\mu\right)_{;\mu} + 8\pi J^\mu_{;\mu} + 4\left(\beta^2 \Gamma^\nu_{[\underline{\mu\nu}]}\right)_{;\mu} = 0 , \tag{29}$$

and

$$3\left(\beta^2 W^\mu\right)_{;\mu} + 4\pi \Omega^{[\mu\nu]}_{\nu;\mu} - 2\left(\beta^2 \Gamma^\nu_{[\underline{\mu\nu}]}\right)_{;\mu} = 0 . \tag{30}$$

Comparing between (23) and (28) one can obtain

$$2\Gamma^\alpha_{[\underline{\mu\nu}];\alpha;\nu} = -(\tfrac{1}{2})k\beta^2 W^\mu - 4\pi \left(J^\mu + \Omega^{[\mu\nu]}_\nu\right) . \tag{31}$$

Expression (31) may be generalized as

$$2\Gamma^\alpha_{[\underline{\mu\nu}];\alpha;\lambda} = -(\tfrac{1}{6})k\beta^2\left(\delta^\nu_\lambda W^\mu - \delta^\mu_\lambda W^\nu\right) - (\tfrac{4\pi}{3})\left(J^\mu\delta^\nu_\lambda - J^\nu\delta^\mu_\lambda + 3\Omega^{[\mu\nu]}_\lambda\right) . \tag{32}$$

For a moment let us go back to (23). The form of this equation justifies introducing the following strength tensor of the electromagnetic field:

$$\Phi_{\mu\nu} = W_{\mu\nu} - 2\Gamma^\alpha_{[\mu\nu];\alpha} \equiv W_{\mu;\nu} - W_{\nu;\mu} - 2\Gamma^\alpha_{[\mu\nu];\alpha} . \tag{33}$$

With (33) one can rewrite the field equation (23) as

$$\Phi^{\mu\nu}_{;\nu} = (\tfrac{1}{2})\beta^2(k-6)W^\mu + 2\beta^2\Gamma^\alpha_{[\underline{\mu\alpha}]} + 4\pi J^\mu . \tag{34}$$

In order to get the equations for the dual field tensor, we make use of (32), and (33), and write down the cyclic permutation:

$$\Phi_{\mu\nu;\lambda} + \Phi_{\lambda\mu;\nu} + \Phi_{\nu\lambda;\mu} = 4\pi\left(\Omega_{\mu[\nu\lambda]} + \Omega_{\lambda[\mu\nu]} + \Omega_{\nu[\lambda\mu]}\right) \equiv 4\pi\,\Theta_{\mu\nu\lambda} , \tag{35}$$

with $\Omega_{\lambda[\mu\nu]} \equiv g_{\mu\alpha}g_{\nu\beta}\Omega_\lambda^{[\alpha\beta]}$ (cf. (27)).

The dual field tensor may be introduced in the usual way (cf. [26])

$$\tilde{\Phi}^{\mu\nu} = -\frac{1}{2\sqrt{-g}}\varepsilon^{\mu\nu\alpha\beta}\Phi_{\alpha\beta} \tag{36}$$

where $\varepsilon^{\mu\nu\alpha\beta}$ stands for the completely antisymmetric Levi-Civita symbol, and $\varepsilon^{0123} = 1$.

Making use of (36), and (35) we obtain the following equation for the dual field

$$\tilde{\Phi}^{\mu\nu}_{;\nu} = 4\pi\,L^\mu , \tag{37}$$

with the dual current given by

$$L^\mu = -\frac{1}{6\sqrt{-g}} \varepsilon^{\mu\lambda\nu\sigma} \Theta_{\lambda\nu\sigma}. \tag{38}$$

$L^\mu$ will be regarded as the intrinsic magnetic current density vector. From (37) we have the conservation law

$$L^\mu_{;\mu} = 0 . \tag{39}$$

We obtained two equations (34), and (37) for the field strength tensor $\Phi_{\mu\nu}$, and its dual $\tilde\Phi_{\mu\nu}$. These fields are caused by intrinsic magnetic currents, electric currents, as well (for $k \neq 6$) by a Proca term.

Further, varying in (22) the metric tensor $g_{\mu\nu}$ one obtains the gravitational field equation

$$\beta^2 G^{\mu\nu} = -8\pi T^{\mu\nu} - 8\pi\left(\tilde{M}^{\mu\nu} - \overline{M}^{\mu\nu}\right) + 2\beta\left(g^{\mu\nu}\beta_{;\alpha;\underline{\alpha}} - \beta_{;\mu;\underline{\nu}}\right) + 4\beta_{,\mu}\beta_{,\nu} - g^{\mu\nu}\beta_{,\alpha}\beta_{,\alpha}$$
$$+ (k-6)\beta^2\left(W^\mu W^\nu - \tfrac{1}{2} g^{\mu\nu} W^\sigma W_\sigma\right) + 4\beta^2 \Gamma^\alpha_{[\sigma\alpha]}\left(g^{\sigma\nu} W^\mu + g^{\sigma\mu} W^\nu - g^{\mu\nu} W^\sigma\right)$$
$$+ \beta^2 \Gamma^\alpha_{[\sigma\tau]}\Gamma^\omega_{[\lambda\rho]} f(g,\delta) - g^{\mu\nu}\beta^4\Lambda , \tag{40}$$

where the expression $f(g,\delta)$ is given by

$$f(g,\delta) = 2\delta^\tau_\alpha \delta^\rho_\omega \left(g^{\lambda\sigma} g^{\mu\nu} - g^{\lambda\mu} g^{\sigma\nu} - g^{\lambda\nu} g^{\sigma\mu}\right) + \delta^\tau_\omega \delta^\rho_\alpha \left(g^{\lambda\mu} g^{\sigma\nu} + g^{\lambda\nu} g^{\sigma\mu} - g^{\lambda\sigma} g^{\mu\nu}\right)$$
$$+ g^{\tau\rho}\left(2 g^{\lambda\mu} g^{\sigma\nu} g_{\alpha\omega} - \delta^\mu_\omega \delta^\nu_\alpha g^{\lambda\sigma} - \tfrac{1}{2} g^{\mu\nu} g^{\lambda\sigma} g_{\alpha\omega}\right). \tag{41}$$

In equation (40) $G^{\mu\nu}$ stands for the Einstein tensor formed with the Christoffel symbols $\{^\lambda_{\mu\nu}\}$, further $\Lambda$ is the cosmological constant, the energy-momentum density tensor of matter is given by $8\pi T^{\mu\nu} = \delta L_m/\delta g_{\mu\nu}$, and the modified energy-momentum density tensors of the field are introduced as follows

$$4\pi \tilde{M}^{\mu\nu} = \tfrac{1}{4} g^{\mu\nu} \Phi^{\alpha\beta}\Phi_{\alpha\beta} - \Phi^{\mu\alpha}\Phi^\nu_{\ \alpha} , \tag{42}$$

and

$$4\pi \overline{M} = \tfrac{1}{4} g^{\mu\nu}\left(\Phi^{\alpha\beta} - W^{\alpha\beta}\right)\left(\Phi_{\alpha\beta} - W_{\alpha\beta}\right) - \left(\Phi^{\mu\alpha} - W^{\mu\alpha}\right)\left(\Phi^\nu_{\ \alpha} - W^\nu_{\ \alpha}\right) . \tag{43}$$

Varying in (22) the gauge function $\beta$ one obtains the following equation:

$$\beta R + k\beta_{;\underline\alpha;\alpha} = \beta\left(8\Gamma^\alpha_{[\lambda\alpha]} w^\lambda - 4\Gamma^\alpha_{[\underline\lambda\alpha];\lambda} - 4\Gamma^\alpha_{[\lambda\alpha]}\Gamma^\rho_{[\underline\lambda\rho]} + 2\Gamma^\alpha_{[\lambda\rho]}\Gamma^\rho_{[\underline\lambda\alpha]} + g_{\alpha\omega}\Gamma^\alpha_{[\lambda\rho]}\Gamma^\omega_{[\underline\lambda\rho]}\right)$$
$$+ \beta(k-6)\left(w^\sigma w_\sigma - w^\sigma_{;\sigma}\right) + 4\beta^3\Lambda + 8\pi B. \tag{44}$$

where $R$ stands for the Riemannian curvature scalar, and $16\pi B = \delta L_m/\delta\beta$ .

Contracting (40), and comparing with (44), one obtains

$$(k-6)\left(\beta^2 W^\lambda\right)_{;\lambda} + 8\pi\left(T - \beta B\right) + 4\left(\beta^2 \Gamma^{\sigma}_{[\lambda\sigma]}\right)_{;\lambda} = 0 . \tag{45}$$

Comparing (45) with (29) one gets the following simple relation:

$$J^\lambda_{;\lambda} = T - \beta B \tag{46}$$

Finally, considering for the in-scalar $\int L_m \sqrt{-g}\, d^4x$ an infinitesimal transformation of coordinates one can derive the following energy-momentum conservation law:

$$T^\lambda_{\mu;\lambda} + J^\lambda_{;\lambda} w_\mu + J^\lambda W_{\mu\lambda} - B\beta_{;\mu} - \Omega^{[\sigma\rho]}_\lambda \Gamma^\lambda_{[\sigma\rho];\mu} + \left(2\Omega^{[\rho\sigma]}_\lambda \Gamma^\lambda_{[\rho\mu]} - \Omega^{[\rho\lambda]}_\mu \Gamma^\sigma_{[\rho\lambda]}\right)_{;\sigma} = 0 . \tag{47}$$

Considering for the same matter integral an infinitesimal gauge transformation (cf.(13) - (15)) we obtain again relation (46).

The equations derived in this section are covariant under general coordinate transformations, as well under gauge transformations (13) - (15). The electric current density $J^\mu$ is introduced as the variational derivative of the matter Lagrangian density with respect to the Weyl connection vector (cf.(25)), while the magnetic current density $L^\mu$ is expressed in terms of the variational derivative with respect to the torsion tensor $\Gamma^\lambda_{[\mu\nu]}$ (cf. (27), (35), (38)).

## 5. A MASSIVE ELECTRODYNAMICS WITH MAGNETIC CURRENTS

Above we obtained the framework for a theory of electromagnetism, that contains intrinsic magnetic, and electric currents. In this framework the torsion plays a crucial role. With it one can generate a dual field tensor having a non-vanishing divergence, so that finally magnetic currents become possible. Generally the torsion tensor has 24 components and it may be chosen in various ways. In order to have magnetic currents in addition to the electric ones, created by the vector $w_\mu$, we need an additional creator vector, which we will build from the torsion. This may be broken in three irreducible parts (cf. e.g. [27],[28]): a trace part, a traceless one, and a totally antisymmetric part. For our purpose only the third part is essential, so that with two auxiliary tensors

$$\Gamma_{\lambda[\mu\nu]} = g_{\lambda\sigma}\Gamma^\sigma_{[\mu\nu]} , \qquad \Gamma^{\lambda[\mu\nu]} = g^{\mu\rho}g^{\nu\sigma}\Gamma^\lambda_{[\rho\sigma]} ; \tag{48}$$

we can write

$$\Gamma_{\lambda[\mu\nu]} = \sqrt{-g}\, \varepsilon_{\lambda\mu\nu\sigma} V^\sigma , \qquad \Gamma^{\lambda[\mu\nu]} = -\frac{1}{\sqrt{-g}}\, \varepsilon^{\lambda\mu\nu\sigma} V_\sigma . \tag{49}$$

Comparing (48), and (49) with the relations (13) - (15) we conclude that $V_\mu$ is a gauge-invariant vector. From (48), (49) we also have the useful formulae

$$\Gamma^{\sigma}_{[\mu\sigma]} = 0, \quad \Gamma^{\lambda[\mu\nu]}_{;\lambda} = \frac{\varepsilon^{\mu\nu\alpha\sigma}}{2\sqrt{-g}}\left(V_{\alpha;\sigma} - V_{\sigma;\alpha}\right). \tag{50}$$

Making use of (48) - (50) one can rewrite the field strength tensor defined by (33) as

$$\Phi^{\mu\nu} = \left(W^{\mu}_{;\underline{\nu}} - W^{\nu}_{;\underline{\mu}}\right) - \frac{\varepsilon^{\mu\nu\alpha\sigma}}{\sqrt{-g}}\left(V_{\alpha;\sigma} - V_{\sigma;\alpha}\right), \tag{51}$$

and by (36) its dual may be written as

$$\tilde{\Phi}^{\mu\nu} = -2\left(V^{\mu}_{;\underline{\nu}} - V^{\nu}_{;\underline{\mu}}\right) - \frac{\varepsilon^{\mu\nu\alpha\sigma}}{2\sqrt{-g}}\left(W_{\alpha;\sigma} - W_{\sigma;\alpha}\right). \tag{52}$$

Substituting (50) - (52) into (34), and (37) one obtains the field equations

$$\Phi^{\mu\nu}_{;\nu} = W^{\mu}_{;\underline{\nu};\nu} - W^{\nu}_{;\underline{\mu};\nu} = \left(\tfrac{1}{2}\right)(k-6)\beta^{2}W^{\mu} + 4\pi J^{\mu}, \tag{53}$$

and

$$\tilde{\Phi}^{\mu\nu}_{;\nu} = V^{\mu}_{;\underline{\nu};\nu} - V^{\nu}_{;\underline{\mu};\nu} = -2\pi L^{\mu}. \tag{54}$$

Now, let us consider $\Omega^{[\mu\nu]}_{\lambda}$. Comparing (53) with (28), we obtain

$$\Omega^{[\mu\nu]}_{\nu} = J^{\mu} + \left(\frac{1}{8\pi}\right)\beta^{2}kW^{\mu}. \tag{55}$$

On the other hand the symmetry properties of $\Omega^{[\mu\nu]}_{\lambda}$, given by (27), must be in accordance with these of $\Gamma^{\lambda}_{[\mu\nu]}$. An appropriate choice is:

$$\Omega^{[\mu\nu]}_{\lambda} = \tfrac{1}{3}\left(\delta^{\nu}_{\lambda}J^{\mu} - \delta^{\mu}_{\lambda}J^{\nu}\right) + \frac{k\beta^{2}}{24\pi}\left(\delta^{\nu}_{\lambda}W^{\mu} - \delta^{\mu}_{\lambda}W^{\nu}\right) - \frac{g_{\lambda\rho}}{\sqrt{-g}}\varepsilon^{\rho\mu\nu\sigma}l_{\sigma}. \tag{56}$$

Making use of (35), (38) and (56) we obtain

$$L_{\mu} = 3l_{\mu}. \tag{57}$$

and from (53), and (54) we can write down the current conservation laws

$$(k-6)\left(\beta^{2}W^{\lambda}\right)_{;\lambda} + 8\pi J^{\lambda}_{;\lambda} = 0, \quad L^{\lambda}_{;\lambda} = 0. \tag{58}$$

Finally, making use of (48) - (50), and discarding the cosmological term $(\beta^{4}\Lambda)$ we can rewrite (40) as

$$G^{\mu\nu} = -(8\pi/\beta^2)(T^{\mu\nu} + \tilde{M}^{\mu\nu} - \overline{M}^{\mu\nu}) + (2/\beta)\left(g^{\mu\nu}\beta_{;\alpha;\alpha} - \beta_{;\mu;\nu}\right)$$
$$+ (1/\beta^2)\left(4\beta_{;\mu}\beta_{;\nu} - g^{\mu\nu}\beta_{;\alpha}\beta_{;\alpha}\right) + (k-6)\left(W^\mu W^\nu - \tfrac{1}{2}g^{\mu\nu}W^\sigma W_\sigma\right) - g^{\mu\nu}V^\sigma V_\sigma - 2V^\mu V^\nu. \tag{59}$$

Equations (53), and (54) describe electromagnetic fields, characterized by two vectors: $W_\mu$, and $V_\mu$, and induced by electric and magnetic currents. It is worth noting that (53) is rather a Proca equation then a Maxwell one. The influence of the electromagnetic field, on the gravitational field is manifested in eq. (59). Making use of the contracted Bianchi identity one obtains from (59) the energy-momentum conservation law. The equations, as well as the conservation laws of this massive torsional electrodynamics are covariant under gauge transformations (13) - (15).

Let us consider the theory in the Einstein gauge ( cf. [21] ). Setting

$$\beta = 1 \tag{60}$$

we replace $W_\mu$ by $w_\mu$ (cf.(24)), and obtain from (51), (52) the field and its dual

$$\Phi^{\mu\nu} = w^\mu_{;\nu} - w^\nu_{;\mu} - \frac{\varepsilon^{\mu\nu\lambda\sigma}}{\sqrt{-g}}\left(V_{\lambda;\sigma} - V_{\sigma;\lambda}\right), \tag{61}$$

and

$$\tilde{\Phi}^{\mu\nu} = -2\left(V^\mu_{;\nu} - V^\nu_{;\mu}\right) - \frac{\varepsilon^{\mu\nu\lambda\sigma}}{2\sqrt{-g}}\left(w_{\lambda;\sigma} - w_{\sigma;\lambda}\right). \tag{62}$$

The field equation (54) remains unchanged, while (53) now may be written as

$$\Phi^{\mu\nu}_{;\nu} = w^\mu_{;\nu;\nu} - w^\nu_{;\mu;\nu} = -\kappa^2 w^\mu + 4\pi J^\mu, \tag{63}$$

where $k$ is replaced by a new parameter $\kappa^2 = (1/2)(6-k)$. Equation (59) takes on the form:

$$G^{\mu\nu} = -8\pi\left(T^{\mu\nu} + \tilde{M}^{\mu\nu} - \overline{M}^{\mu\nu}\right) - 2\kappa^2\left(w^\mu w^\nu - \tfrac{1}{2}g^{\mu\nu}w^\lambda w_\lambda\right) - 2V^\mu V^\nu - g^{\mu\nu}V^\lambda V_\lambda. \tag{64}$$

The current conservation laws (58) now are written as

$$\left(4\pi J^\lambda - \kappa^2 w^\lambda\right)_{;\lambda} = 0, \quad \text{and} \quad L^\lambda_{;\lambda} = 0. \tag{65}$$

For a moment let us consider a current free ( $J^\mu = 0$ ) small region. For $\kappa \neq 0$ there is holding the condition

$$w^\lambda_{;\lambda} = 0, \tag{66}$$

which is reminiscent of the Lorentzian gauge conditions in ordinary electrodynamics. Making use of (66) one can rewrite (63) as

$$w^\mu_{;\nu;\nu} + w^\nu R^\mu_\nu + \kappa^2 w^\mu = 0, \tag{67}$$

with $R^\mu_\nu$ being the Ricci tensor formed from the Christoffel symbols $\{^\lambda_{\sigma\rho}\}$. If the curvature in the current-free region is negligible, one has the Proca equation [25] for a vector field $w^\mu$

$$w^\mu_{;\,\nu;\,\nu} + \kappa^2 w^\mu = 0, \tag{68}$$

which from the quantum mechanical point of view describes a particle with spin 1, and mass given in conventional units by

$$m_\gamma = \left(\frac{h}{c}\right)\kappa \equiv \left(\frac{h}{c}\right)\sqrt{\frac{6-k}{2}} \ . \tag{69}$$

Hence for $k < 6$ one has massive field particles, photons. Below it will be shown that these massive photons are essential in order to ensure interaction between magnetic currents.

Let us turn to the energy-momentum conservation law in the Einstein gauge. This law may be obtained by complicated and cumbersome calculations from (47), if one makes use of (42), (43), (54), (60)-(63). However there is a more convenient procedure, based on the contracted Bianchi identity

$$G^\nu_{\mu;\,\nu} = 0 \tag{70}$$

Making use of (70) we obtain from (64)

$$8\pi\left(T^\nu_{\mu;\nu} + \tilde{M}^\nu_{\mu;\nu} - \overline{M}^\nu_{\mu;\nu}\right) + 2\kappa^2\left(w_\mu w^\nu - \tfrac{1}{2}\delta^\nu_\mu w_\lambda w^\lambda\right)_{;\nu} + 2(V_\mu V^\nu)_{;\nu} + (V_\lambda V^\lambda)_{;\mu} = 0. \tag{71}$$

Taking into account the definitions (42), and (43), as well the field equations (54), (63), we can rewrite the energy-momentum conservation law (71) as follows:

$$4\pi T^\nu_{\mu;\nu} + 4\pi \Phi_{\mu\sigma} J^\sigma + 2\pi\sqrt{-g}\,\varepsilon_{\alpha\beta\mu\sigma}W^{\alpha\beta}L^\sigma - \kappa^2\left(\Phi_{\mu\sigma} + W_{\mu\sigma}\right)w^\sigma \\ + \kappa^2 w_\mu w^\nu_{;\nu} + V^\nu\left(V_{\mu;\nu} + V_{\nu;\mu}\right) + V_\mu V^\nu_{;\nu} = 0. \tag{72}$$

Let us consider a region free of matter, and currents, where
$$T^\nu_\mu = 0\,,\qquad J^\nu = 0\,,\qquad L^\nu = 0\ . \tag{73}$$

As in this region condition (66) is holding, one has from (72)

$$\kappa^2\left(\Phi_{\mu\sigma} + W_{\mu\sigma}\right)w^\sigma + V^\nu\left(V_{\mu;\nu} + V_{\nu;\mu}\right) + V_\mu V^\nu_{;\nu} = 0. \tag{74}$$

From (74) we can conclude that in absence of magnetic fields $V^\mu$, invoked by intrinsic magnetic currents $L^\mu$ (cf. (54)), we have $\kappa = 0$. Thus the photon is massless in absent of magnetic fields. On the other hand the magnetic field vector $V^\mu$, may be accompanied by either massless, or massive photons.

From (72) one can derive the equation of motion of a test particle. Suppose we have a field in vacuum, so that (66), and (74) are holding. In a given field ($\Phi^{\mu\nu}$, $W^{\mu\nu}$) let us now consider matter in the form of pressureless dust consisting of identical particles, each having mass (rest energy) $m_0$, electric charge $\varepsilon_0$, and four velocity $u^\mu$. Then we can rewrite (72) as

$$4\pi T^{\nu}_{\mu;\nu} + 4\pi \Phi_{\mu\sigma} J^\sigma + 2\pi \sqrt{-g}\, \varepsilon_{\alpha\beta\mu\sigma} W^{\alpha\beta} L^\sigma = 0 , \qquad (75)$$

where the energy-momentum density tensor is given by

$$T^{\nu}_{\mu} = \rho\, u_\mu u^\nu , \qquad (76)$$

and the mass density $\rho$ is

$$\rho = m_0\, n , \qquad (77)$$

with $n$ being the particle density. The conservation of the number of particles is described by

$$\left(n u^\lambda\right)_{;\lambda} = 0. \qquad (78)$$

Further the electric current density vector may be written as

$$J^\mu = \varepsilon_0\, n u^\mu \equiv \left(\frac{\varepsilon_0}{m_0}\right) \rho\, u^\mu . \qquad (79)$$

Setting $L^\mu = 0$, and making use of (76)-(79), we get from (75) the equation of motion

$$\frac{du^\mu}{ds} + \left\{{}^{\mu}_{\lambda\sigma}\right\} u^\lambda u^\sigma + \frac{\varepsilon_0}{m_0} u_\lambda \Phi^{\mu\lambda} = 0. \qquad (80)$$

In order to consider a test particle having mass $m_0$, and magnetic charge $\mu_0$, we set $J^\mu = 0$, and write the magnetic current density vector as

$$L^\nu = \mu_0\, n u^\nu \equiv \left(\frac{\mu_0}{m_0}\right) \rho\, u^\nu . \qquad (81)$$

Then from (75) we obtain

$$\frac{du^\nu}{ds} + \left\{{}^{\nu}_{\lambda\sigma}\right\} u^\lambda u^\sigma + \frac{1}{2}\frac{\mu_0}{m_0} u^\sigma \sqrt{-g}\, \varepsilon_{\alpha\beta\lambda\sigma} g^{\lambda\nu} W^{\alpha\beta} = 0, \qquad (82)$$

where the field $W^{\alpha\beta}$ is given by (cf. (12))

$$W^{\alpha\beta} = w^{\alpha}_{;\beta} - w^{\beta}_{;\alpha} \qquad (83)$$

Let us imagine an electrically charged test particle moving in the field of massive bodies carrying electric and magnetic charges. The equation of motion, (80) contains the field strength tensor $\Phi^{\mu\nu}$ depending on two field vectors (cf.(61)) $w^\mu$, and $V^\mu$. But according to (63) $w^\mu$ may be created by electric charges, as well by a self-inducing Proca term, while according to (54) $V^\mu$ is created by magnetic charges. Thus on the electrically charged test particle act electric, as well magnetic field sources, and the photon in this case may be either massive, or massless.

Now, let us turn to the magnetically charged test particle. In equation (82) the field is represented by $W^{\alpha\beta}$, and according to (83), and (63) one concludes that the test particle is affected by electrically charged bodies and by massive photons. Thus, where is the interaction between two magnetic monopoles? The answer is given by equation (74). From it we see that the magnetic vector $V^\mu$ may be accompanied by a massive photon, the latter contributing to the creation of $W^{\alpha\beta}$. Thus, two magnetically charged bodies interact by means of massive photons.

## 6. SPHERICAL SYMMETRY

Let us choose the Einstein gauge $\beta = 1$ (cf. (60)). Suppose there is a particle at rest in the origin, and we consider a region around the particle, which is so small that we can neglect the cosmic curvature. In this case the static spherically symmetric line-element may be written as

$$ds^2 = e^\nu dt^2 - e^\lambda dr^2 - r^2\left(d\vartheta^2 + \sin^2\vartheta \, d\varphi^2\right), \tag{84}$$

with $\nu$, and $\lambda$ being functions of $r$. If the particle is charged it may create fields $w^\mu$, and $V^\mu$. From symmetry reasons there is only one non-vanishing component of each vector

$$w_0 \equiv w(r), \quad \text{and} \quad V_0 \equiv V(r). \tag{85}$$

Making use of (42), (43), (60)-(62), as well of the metric (84) we obtain from (64) the following non-zero Einstein equations explicitly:
for $\mu = \nu = 0$,

$$e^{-\lambda}\left(-\frac{\lambda'}{r} + \frac{1}{r^2}\right) - \frac{1}{r^2} = -8\pi T_0^0 - e^{-(\lambda+\nu)}(w')^2 - \kappa^2 e^{-\nu} w^2 - 3e^{-\nu}V^2, \tag{86}$$

for $\mu = \nu = 1$,

$$e^{-\lambda}\left(\frac{\nu'}{r} + \frac{1}{r^2}\right) - \frac{1}{r^2} = -8\pi T_1^1 - e^{-(\lambda+\nu)}(w')^2 + \kappa^2 e^{-\nu} w^2 - e^{-\nu}V^2, \tag{87}$$

and for $\mu = \nu = 2$ (or for $\mu = \nu = 3$),

$$e^{-\lambda}\left[\nu'' + \tfrac{1}{2}(\nu')^2 + \frac{1}{r}(\nu' - \lambda') - \tfrac{1}{2}\lambda'\nu'\right] = -16\pi T_2^2 + 2e^{-(\lambda+\nu)}(w')^2 + 2\kappa^2 e^{-\nu} w^2 - 2e^{-\nu}V^2, \tag{88}$$

with $f'$ standing for $df/dr$.

The equations for the two vectors $w^\mu$, and $V^\mu$, (54), and (63) in this case take on the form

$$w'' - \frac{1}{2}(\lambda' + \nu')w' + \frac{2}{r}w' = \kappa^2 e^\lambda w - 4\pi\, e^\lambda\, J_0\,, \tag{89}$$

and

$$V'' - \frac{1}{2}(\lambda' + \nu')V' + \frac{2}{r}V' = 2\pi\, e^\lambda L_0\,. \tag{90}$$

Integrating (89) one obtains

$$w' = \frac{e^{\frac{1}{2}(\lambda+\nu)}}{r^2}\left[q(r) + \kappa^2 I(r) + Q\right], \tag{91}$$

with $Q$ = const. being the charge located in the origin, with the electric charge within a sphere of radius $r$ given by

$$q(r) = -4\pi\int_0^r e^{\frac{1}{2}(\lambda-\nu)} J_0\, r^2 dr = 4\pi\int_0^r e^{\frac{\lambda}{2}} \rho_e\, r^2 dr. \tag{92}$$

($\rho_e$ is the ordinary three-dimensional charge density), and with the Proca charge

$$I(r) = \int_0^r e^{\frac{1}{2}(\lambda-\nu)} w r^2 dr. \tag{93}$$

From (90) one obtains

$$V' = \frac{e^{\frac{1}{2}(\lambda+\nu)}}{r^2}\left[l(r) + \tilde{M}\right], \tag{94}$$

where $\tilde{M}$ stands for the magnetic charge located in the origin, and $l(r)$ is given by

$$l(r) = 2\pi \int_0^r e^{\frac{1}{2}(\lambda-\nu)} L_0\, r^2 dr\,. \tag{95}$$

Finally, for the given symmetry, one can write down the energy-momentum conservation law (72), stemming from the Bianchi identity, as follows

$$4\pi\left[(T_1^1)' + \frac{1}{2}\nu'(T_1^1 - T_0^0) + \frac{2}{r}(T_1^1 - T_2^2) - J_0 w' e^{-\nu}\right] \\ + \left[-2\kappa^2 w w' + V V' - V^2 \nu'\right]e^{-\nu} = 0. \tag{96}$$

For the above treated spherically symmetric case one has to solve the field equations (91), (94) together with the Einstein equations (86) - (88). If one wants one can replace one of the equations (86) - (88) by the energy condition (96).

One can think about vacuum surrounding the particle, and about massless photons. In this simple case

$$T_\mu^\nu = 0; \quad J^\mu = 0; \quad L^\mu = 0; \quad \kappa = 0. \tag{97}$$

and (91), and (94) may be written as

$$w' = \frac{e^{\frac{1}{2}(\lambda+\nu)}}{r^2} Q, \tag{98}$$

and

$$V' = \frac{e^{\frac{1}{2}(\lambda+\nu)}}{r^2} \tilde{M}. \tag{99}$$

Making use of (97), and (98) we can rewrite (86) in the following form:

$$e^{-\lambda}\left(-\frac{\lambda'}{r}+\frac{1}{r^2}\right) - \frac{1}{r^2} = -\frac{Q^2}{r^4} - 3 e^{-\nu} V^2, \tag{100}$$

and subtracting (86) from (87), and making use of (97), (98) we obtain

$$e^{-\lambda}\left(\frac{\lambda'+\nu'}{r}\right) = 2e^{-\nu}V^2. \tag{101}$$

Further, making use of (97), we obtain from (96)

$$VV' - V^2 \nu' = 0, \tag{102}$$

and integrating this we have

$$V = Ke^\nu, \quad (K = const). \tag{103}$$

Inserting (103) into (101) and integrating we obtain

$$e^{\lambda+\nu} = \left(B^2 - K^2 r^2\right)^{-1}, \tag{104}$$

with $B$ being an arbitrary constant. Let us introduce an auxiliary function

$$y = e^{-\lambda}. \tag{105}$$

Then, making use of (104), and (105), we get from (100) the following equation:

$$\frac{1}{r}y' + \frac{1}{r^2}y + \frac{3K^2 y}{B^2 - K^2 r^2} = \frac{1}{r^2} - \frac{Q^2}{r^4}. \tag{106}$$

The solution of (106) may be written as follows

$$y = e^{-\lambda} = \frac{1}{B}\left(\frac{1}{r^2} - \frac{K^2}{B^2}\right)\left[Br^2 + Q^2\left(B^2 - 2K^2r^2\right)\right] + \frac{K_1\left(B^2 - K^2r^2\right)^{\frac{3}{2}}}{B^3 r} . \quad (107)$$

where $K_1$ is a constant. From (104) we can now obtain

$$e^{\nu} = \frac{1}{B^3 r^2}\left[Br^2 + Q^2\left(B^2 - 2K^2r^2\right)\right] + \frac{K_1\left(B^2 - K^2r^2\right)^{\frac{1}{2}}}{B^3 r} . \quad (108)$$

We recall that $Q$ is the charge located in the origin (cf. (98)), while $K$ represents the magnetic field (cf. (103)). In order to get from (107), (108) the Schwarzschild solution, in absent of electric and magnetic fields ($Q = 0$, $K = 0$.), we must set

$$B = 1, \quad K_1 = -2m , \quad (109)$$

with $m$ being the mass of the particle. In absent of magnetic fields ($K = 0$), and with (109) we have from (107), (108) the Reissner-Nordstroem solution

$$e^{-\lambda} = e^{\nu} = 1 + \frac{Q^2}{r^2} - \frac{2m}{r} . \quad (110)$$

If we want to have the solution describing a magnetic monopole, we must satisfy relations (99), and (103). Taking into account (109), and (108) we obtain

$$V' = \varepsilon K\left(e^{\nu}\right)' = \varepsilon K\left[-\frac{2Q^2}{r^3} + \frac{2m}{r^2\left(1 - K^2r^2\right)^{\frac{1}{2}}}\right], \quad (111)$$

where the unit charge $\varepsilon$ (magnetic, as well electric) is introduced in order to keep the dimensions identical with those of general relativity. From (111), and (104) we see that (99) will be satisfied providing we set $Q = 0$. By this condition we can rewrite (107), and (108) as

$$e^{-\lambda} = \left(1 - K^2r^2\right)\left[1 - \frac{2m}{r}\sqrt{1 - K^2r^2}\right], \quad (112)$$

and

$$e^{\nu} = 1 - \frac{2m}{r}\sqrt{1 - K^2r^2} . \quad (113)$$

Finally, comparing (99) and (111), we get

$$K = \frac{\tilde{M}}{2\varepsilon m} , \quad (114)$$

with $\tilde{M}$ being the magnetic charge in the origin. One can see from (114) that the magnetic monopole is massive. Finally, according to (99), and (111) the field strength is given by

$$\tilde{\Phi}_{01} = V' = \varepsilon\, K(e^v)' = \frac{\tilde{M}}{r^2\sqrt{1-K^2 r^2}}. \tag{115}$$

In (112)-(115) one has the metric and the field of a magnetic monopole located in the origin. This solution is defined in the interval

$$0 < r < r_b = \frac{1}{K} = \frac{2\varepsilon\, m}{\tilde{M}}. \tag{116}$$

According to (112), (113) the monopole is surrounded by a spherical surface of radius

$$r = r_s = \frac{2m}{\sqrt{1+\frac{\tilde{M}^2}{\varepsilon^2}}}, \tag{117}$$

on which one has

$$e^{-\lambda}(r_s) = 0, \quad e^v(r_s) = 0, \quad V(r_s) = 0, \text{ but } V'(r_s) \neq 0. \tag{118}$$

In absent of magnetic charge ($\tilde{M} = 0$) this surface turns into the Schwarzschild sphere.

One can consider a particle having an elementary magnetic charge, given by the Dirac relation [2] $\tilde{M} = \frac{137}{2} e$ ( $e$ - the electron charge ), and a Planckian mass $m = m_{Pl}$, so that in general relativistic units one has $\tilde{M} \cong 9.1 \times 10^{-33}$ cm, $m \cong 1.608 \times 10^{-33}$ cm. According to (116), and (117) this gives $r_s = 3.22 \times 10^{-33}$ cm, $r_b = 3.53 \times 10^{-1}$ cm.

The line-element that corresponds to (112), (113) has the following form:

$$ds^2 = \left(1 - \frac{2m}{r}\sqrt{1-K^2 r^2}\right)dt^2 - \left[(1-K^2 r^2)\left(1-\frac{2m}{r}\sqrt{1-K^2 r^2}\right)\right]^{-1} dr^2 - r^2 d\Omega^2, \tag{119}$$

and it is defined for $0 \le r \le r_b = K^{-1}$. If one wants, he can turn to a metric with a new radial variable $R$ given by

$$R^2 = \frac{r^2}{1-K^2 r^2}; \quad \Rightarrow \quad r^2 = \frac{R^2}{1+K^2 R^2}, \tag{120}$$

so that the new range is

$$0 \le R < \infty. \tag{121}$$

For the new line-element one obtains

$$ds^2 = \left(1 - \frac{2m}{R}\right)dt^2 - \left[\left(1 + K^2 R^2\right)^2 \left(1 - \frac{2m}{R}\right)\right]^{-1} dR^2 - \frac{R^2}{1 + K^2 R^2} d\Omega^2 , \qquad (122)$$

while the strength of the magnetic field is now given by the very simple relation

$$\tilde{\Phi}_{01} = \frac{\tilde{M}}{R^2} . \qquad (123)$$

The surface of metric singularity in the new representation is exactly the Schwarzschild sphere with

$$R_s = 2m. \qquad (124)$$

Thus, in the framework of the Weyl-Dirac torsional electrodynamics the metric, and magnetic field strength of a magnetic monopole are given by equations (119), and (115) , or alternatively by (122), (123), the monopole is surrounded by a surface of metric singularity (cf.(117), or (124)), it must be massive, and it can not be located together with an electric monopole. However, it must be noted that the solution was obtained on condition that $\kappa = 0$ (cf. (97)), so that the interaction between magnetic charges was excluded from the scenario (cf. the discussion after formula (83)).

Below we show how a generalized scenario, with a massive photon may be built up. Suppose we have $T_\mu^\nu = 0;\ \ J^\mu = 0;\ \ L^\mu = 0;\ \ Q = 0;$ but $\kappa \neq 0$ . Then eq. (99) is holding, and instead of (98) we obtain from (91)

$$w' = \kappa^2 \frac{e^{\frac{1}{2}(\lambda+\nu)}}{r^2} I(r) = \kappa^2 \frac{e^{\frac{1}{2}(\lambda+\nu)}}{r^2} \int_0^r e^{\frac{1}{2}(\lambda-\nu)} w r^2 dr , \qquad (125)$$

so that instead of (100) we have from (87) the following equation :

$$e^{-\lambda}\left(-\frac{\lambda'}{r} + \frac{1}{r^2}\right) - \frac{1}{r^2} = -\kappa^4 \frac{I^2}{r^4} - \kappa^2 e^{-\nu} w^2 - 3e^{-\nu} V^2, \qquad (126)$$

In the same manner we obtain instead of (101) the equation

$$e^{-\lambda}\left(\frac{\lambda' + \nu'}{r}\right) = 2e^{-\nu}\left[V^2 + \kappa^2 w^2\right], \qquad (127)$$

Finally, from the energy relation (96) we obtain in the present case

$$-2\kappa^2 ww' + VV' - V^2 \nu' = 0. \qquad (128)$$

Thus for the case of a magnetic monopole accompanied by a Proca field we have a complete system of equations. This system includes two Einstein equations, (126), and (127), two field equations, (99), and (125), and the energy relation (128). Explicit solutions of this system will be considered in a subsequent paper.

## 7. DISCUSSION

The Weyl geometry[18] is doubtless the most aesthetic generalization of the Riemannian geometry, the last being the framework of general relativity. In order to build up an action integral, which is coordinate invariant, as well gauge invariant, and which agrees with Einstein's general relativity theory[19], Dirac[20] modified the Weyl theory by introducing a scalar gauge function $\beta$. The Weyl-Dirac theory was discussed, and developed by Rosen[21], who gave a deep analysis of its physical meaning, included a matter term in the action, and considered the outcome of the theory in several gauges, as for instance the cosmic gauge, and the Einstein gauge. Rosen also pointed out that for $k \neq 6$ ($k$ is the Dirac parameter) one has to replace Maxwell equations by Proca ones, and that for $k < 6$ one has massive field quanta.

In the present work the Weyl-Dirac-Rosen framework is generalized. The torsionless Weyl geometry is replaced by a space having both, torsion and Weylian nonmetricity. On this broadened basis we have built up a geometrically based action integral, from which we obtained the field equations, and conservation laws. The outcome is a theory, in which the dual of the field strength tensor has a non-vanishing divergence, and in the equations of the electromagnetic field Proca terms are admitted. Assuming that the torsion has a totally antisymmetric structure, we expressed it in terms of a gauge-invariant vector. The result of our procedure is the torsional Weyl-Dirac electrodynamics (cf. section 5.). This theory possessing intrinsic magnetic, as well electric currents, and admitting massive photons, is covariant under both, coordinate transformations, and the Weyl gauge transformations (cf. (13)-(15)).

In this framework an intrinsic magnetic field generates massive photons, while in absence of magnetic fields ( $V^{\mu} = 0$ ), the photon is massless (cf.(74)). In the latter case one can carry out Maxwellian gauge transformations in addition to the Weyl gauging, and turning to the Einstein gauge one obtains from (63) the ordinary Maxwell equation, from (64) the Einstein equation, and from (65) the electric current conservation law. Thus the Einstein-Maxwell theory is a limiting case of the torsional Weyl-Dirac electrodynamics considered in this work.

The Weyl-Dirac torsional electrodynamics was investigated in the Einstein gauge. It turns out that on electrically charged test particles act electric, as well magnetic currents, and that the electric-electric, and magnetic-electric interactions may be transmitted either by massive, or by massless photons. But considering a magnetically charged test particle we found that the magnetic-magnetic interaction is transmitted by a massive photon. Thus, on one hand the photon is massive only in presence of magnetic fields, and on the other hand the massive photon is responsible for the interaction between magnetic monopoles. A spherically symmetric solutions is obtained. It may represent either the metric, and the field of a magnetic monopole (115), (119), (122), (123), or that of an electric charge, the Reissner-Nordstroem metric (110). Thus, a magnetic monopole can not be located together with an electric one. From (114), and from the line-element (119), (122) one can conclude that the magnetic monopole is massive, and that it is surrounded by a spherical surface on which the metric is singular. In the limiting case, when the magnetic, and electric charges vanish, the solution takes on the Schwarzschild form.

A problem that faces us in Weyl-type geometries is the nonintegrability of length that as though excludes measuring standards in Weyl-type theories. However, one can overcome this obstacle, if he adopts the standpoint that the geometry in the interior of atoms differs from the geometry describing the exterior. A detailed discussion on this subject may be found in a paper of the present writer [30], as well in a recent paper by Wood, and Papini [31].

The Weyl-Dirac torsional electrodynamics is covariant with respect to the Weyl gauge transformations (13) - (15), and one can readily verify, that all equations, as well all quantities having physical meaning are gauge covariant [2]. Owing to this gauge covariance of the theory the physical conclusions found in the Einstein gauge are holding in any gauge.

---

[2]More about gauge covariance of Weyl-type theories and gauge covariant quantities may be found in the works of Dirac [20], Rosen [21], and Canuto et al. [32].